\documentclass [preprint,showpacs,preprintnumbers,amsmath,amssymb, nofootinbib]{revtex4-1}
\usepackage{epsfig}
\usepackage{graphicx}
\usepackage{bm} 
\usepackage{amsmath}

\newcommand{\be}{\begin{equation}}

\newcommand{\ee}{\end{equation}}
\newcommand{\bea}{\begin{eqnarray}}
\newcommand{\eea}{\end{eqnarray}}

\newcommand{\nn}{\nonumber}

\newcommand{\bi}{\begin{enumerate}}
\newcommand{\ei}{\end{enumerate}}
\newcommand{\bref}[1]{(\ref{#1})}
\newcommand{\A}{\alpha} \newcommand{\B}{\beta} \newcommand{\gam}{\gamma}
 \newcommand{\D}{\delta} 
\newcommand{\ep}{\epsilon}

\newcommand{\lam}{\lambda}        \newcommand{\s}{\sigma}
          \newcommand{\w}{\omega}

\newcommand{\W}{\Omega}

\newcommand{\ba}{\overline }\def\6{\partial}\def\7{\tilde}\def\8{\hat}

\newcommand{\bP}{{\bf P}}\newcommand{\bX}{{\bf X}}\newcommand{\bY}{{\bf Y}}
\newcommand{\bM}{{\bf J}}
\newcommand{\bx}{{\bf x}}

\newcommand{\bB}{\bm \beta} 
\newcommand{\sbB}{\bm \beta}

\newcommand{\bCP}{{\cal P}{\hskip-0.30cm{\cal P}}}
\newcommand{\sbCP}{{\cal P}{\hskip-0.21cm{\cal P}}}
\newcommand{\MN}{{M}}
\def\pa{\partial}

\def\CL{{\cal L}}
\def\CA{{\cal A}}\def\CB{{\cal B}}
\def\CH{{\cal H}}
\def\l{{\ell}}
\def\vs{\vskip 4mm}\def\={{\;=\;}}\def\+{{\;+\;}}

\def\NGC{NGC }
 
\begin{document}
\preprint{ICCUB-11-166\, UB-ECM-PF 11/61}

\title{Schrodinger Equations for Higher Order Non-relativistic Particles and 
N-Galilean Conformal  Symmetry}

\author{Joaquim Gomis}\email{gomis@ecm.ub.es}\affiliation{Departament d'Estructura i Constituents de la Mat\`eria}\affiliation{Departament de F\'isica,
Universitat de Barcelona, Diagonal 647, 08028 Barcelona, Spain and
C.E.R. for Astrophysics, Particle Physics and Cosmology, Barcelona
Spain
}
\author{Kiyoshi Kamimura}\email{kamimura@ph.sci.toho-u.ac.jp}
\affiliation{Department of Physics, Toho University, Miyama, Funabashi, 274-8510, Japan}
\date{\today}

\begin{abstract}
We consider Schr\"odinger equations for a non-relativistic particle obeying $N$+1-th order higher derivative classical equation of motion. These equations are invariant under $N$(odd)-extended Galilean conformal (NGC)  algebras  in general $d$+1 dimensions. In 
 2+1 dimensions, the exotic Schr\"odinger equations are invariant under $N$ (even)-GCA. 

\end{abstract}


\maketitle


\section{Introduction}

The Schr\"odinger equation for a non-relativistic free particle 
\be
(i\pa_t+\frac{1}{2M}\nabla^2) \psi(t,\bf{x})=0
\label{Schro}\ee
is invariant under the  scalar projective representation of the Galilei group 
\cite{bargmann}\cite{levyleblond63}.
In addition the maximal symmetry of \bref{Schro}  is  
the Schr\"odinger algebra   \cite{Niederer:1972zz}\cite{Hagen:1972pd}
that includes extra generators associated with dilatation, expansion (special conformal transformations in one dimension) and a central charge. 
The Schr\"odinger equation is obtained by the canonical quantization  of the non-relativistic free particle   whose action 
$S=\int dt\,\frac{M}{2}(\frac{d\bf{X}}{dt})^2$  is 
also invariant under the Schr\"odinger group.   

In this paper we will generalize above result ($N$=1 case)  and show that the higher order non-relativistic particle model given by the Lagrangian\footnote{The $N$=3 case was considered in \cite{Gauntlett:1990nk} as
an example of a realization of the Galilei algebra with zero mass. The
model was also considered in \cite{Lukierski:2007nh}.}
\bea
\CL_X&=&\frac{\MN}2\,
(\frac{d^{\frac{N+1}{2}}\bX}{dt^{\frac{N+1}{2}}})^2,\qquad N=1,3,5,...,
\label{NderLag}\eea
has the $N$-Galilean conformal (\NGC)  symmetry \cite{negro}\cite{Henkel:1997zz}{\bf{\footnote{ In \cite{Duval:2011mi} it was conjectured that $N$+1-th 
order free equations of motion are \NGC invariant for any N, odd and even.}}}. 
$\MN$ is the constant with the dimension $[\MN]=({\rm mass})^{2-N}$.
Furthermore the corresponding Schr\"odinger equations are projective representation of the \NGC symmetries. 
For even $N$ case we consider only 
 2+1 dimensions where the central extension of the \NGC algebra exists.

The \NGC algebra\cite{negro}\cite{Henkel:1997zz} is the finite dimensional extension of the Galilean algebra
for positive integer $N$. Their commutators are 
\bea
\left[D,H\right]&=&i\,H, \quad 
\left[C,H\right]=2i\,D, \quad
\left[D,C\right]=-i\,C, \label{confal1}\nn\\
\left[H,\bCP_j\right]&=&-i\,j\,\bCP_{j-1}, \quad
\left[D,\bCP_j\right]=-i\,(j-\frac{N}2)\,\bCP_{j}, \nn\\ 
\left[C,\bCP_j\right]&=&-i\,(j-N)\,\bCP_{j+1},\quad
\left[\bM_{ab},\bCP_{j,c}\right]=i\,\D_{c[b}\,\bCP_{j,a]},\nn\\
\left[\bM_{ab},\bM_{cd}\right]&=&i\left(\D_{c[b}\bM_{a]d}-\D_{d[b}\bM_{a]c}\right).
\label{confalJM}\eea
It has subalgebras, 
$(H,D,C)$, one dimensional conformal algebra and 
$(H, \bCP_0, \bCP_1,\bM)$, the {unextended} Galilean algebra
{and the acceleration extended Galilean algebra} \cite{Lukierski:2007nh}
 with higher order accelerations $\bCP_j, (j=2,...,N)$. $N$ is interpreted in terms of the dynamical exponent $z$ under the dilation $D$ of the coordinates as
\be
t\to \lam^{z}\,t,\qquad \bX\to   {\lam}\, \bX,\qquad  z=\frac{2}{N}.
\ee
 Recently it gets much attention on the Galilean conformal symmetry and its extensions especially in condensed matter physics and gravity 
\cite{math-ph/0601028}
\cite{arXiv:0810.1545} \cite{arXiv:0902.1385} \cite{Duval:2011mi}. It is interesting to see how such symmetries are realized in a simple particle models.
\vs

In sect.2 we introduce a particle action invariant under central extension of the NGC algebra \bref{confalJM}. In sect.3 we discuss it in quantum theory that how the symmetry is realized in the Schr\"odinger equation.  
Even $N$ cases in $2+1$ dimensions are briefly commented in sect.4 and summary and discussions are in section 5. In appendix we add how the invariant actions are derived using the method of non-linear realization. 

\section{NGC invariant Particle Model}

In order to quantize the particle model described by the Lagrangian 
\bref{NderLag} we construct the Hamiltonian associated with it.
{ Since the theory is higher order in time derivative we 
introduce auxiliary coordinates 
$\bX^j=\dot\bX^{j-1}, \;(1\leq j\leq \frac{N-1}2)$, $\bX^0\equiv\bX$ 
and the Lagrange multipliers $\bY_j$ to get the Lagrangian as}
\bea
\CL_X&=&\frac{\MN}{2}
(\dot\bX^{\frac{N-1}{2}})^2 + \sum_{j=0}^{\frac{N-3}2} \bY_j(\dot\bX^j-\bX^{j+1}).
\label{NderLag1}\eea
The Ostrogradsky momenta \cite{o}  are 
\be
\bP_j=\bY_j, \quad
 (j=0,...,\frac{N-3}2),\quad \bP_{\frac{N-1}{2}}=\MN\,
\dot\bX^{\frac{N-1}{2}}.
\ee
If we use the second class constraint $(\bY_j,\bP^j_Y)=(\bP_j,0)$,
the independent canonical pairs are   $(\bX^j,\bP_j), \;(0\leq j\leq \frac{N-1}2)$ and the Hamiltonian becomes 
\bea\label{hamiltonian}
\CH=\, \sum_{j=0}^{\frac{N-3}2} \bP_j\bX^{j+1}+
\frac1{2\MN}\,{(\bP_{{\frac{N-1}2}})^2}.
\eea
Using the canonical commutators 
\be
[\bX^j ,\bP_k]=i\,\D_{j,k},\qquad [\bX^j ,\bX^k]=[\bP_j ,\bP_k]=0,
\label{Heisenberg}\ee
the Heisenberg equations 
\bea
\dot\bX^j&=&\bX^{j+1},\qquad  \dot\bP_{j+1}=-\bP_{j},\qquad    ( 0\leq j\leq {\frac{N-3}2} ),
\nn\\&&
\dot\bX^{\frac{N-1}2}=\frac{1}{\MN}\bP_{{\frac{N-1}2}},
\qquad  \dot\bP_0=0,\eea
reproduce the Euler-Lagrangian  equation of 
\bref{NderLag}, $\frac{d^{N+1}}{dt^{N+1}}\bX^0=0$.  
The Lagrangian \bref{NderLag1} is 
invariant under the \NGC symmetry 
 whose hermitian canonical generators are 
\bea
H&=&-\CH, \qquad 
\nn\\
D&=&-t \,\CH +\sum_{j=0}^{\frac{N-1}2}\,(\frac{N}2-j)\frac{\{\bX^j, \bP_j\}_+}{2},
\nn\\
C&=&-t^2\,\CH+\sum_{j=0}^{\frac{N-1}2}\,
\left((N-2j)\,t\,\frac{\{\bX^j, \bP_j\}_+}{2}\right.
\nn\\&& \left. +(N-j+1)j\,\bX^{j-1}\,\bP_j\right)
\,-\,{{\frac{M}2\,({\frac{N+1}2})^2(\bX^{{\frac{N-1}2}})^2,}}\nn\\
\bCP_j&=&j!\left( \sum_{\l=0}^j\,\frac{t^{j-\l}}{(j-\l)!}\,\bP_\l\right),\qquad (j=0,...,\frac{N-1}2),
\nn\\
\bCP_j&=&j!\left(  \sum_{\l=0}^{\frac{N-1}2}\,\frac{t^{j-\l}}{(j-\l)!}\,\bP_\l
-\MN\sum_{\l=\frac{N+1}2}^j\,  (-)^{\frac{N+1}2+\l}\times \right.
\nn\\&& \qquad \left.
\,\frac{t^{j-\l}}{(j-\l)!}\,\bX^{N-\l}\right),\qquad (j=\frac{N+1}2,...,N),
\nn\\ {\bf J}_{ab}&=&\sum_{j=0}^{\frac{N-1}2}\,
(\bX^j_{b}\,\bP_{ja}-\bX^j_{a}\,\bP_{jb}).
\label{Canogene1}\eea 
where $\{\bX^j, \bP_k\}_+=\bX^j\bP_k+\bP_k\bX^j$.
These generators, $G$, that have explicit $t$ dependence, are constant of motion, i.e., $\frac{d}{dt}G=-i[G,\CH]+\frac{\pa G}{\pa t}=0$.
It proves the transformations generated by $G $'s are symmetry of the Lagrangian \bref{NderLag1}. Transformations of $\bY_j$ are given by those of $\bP_j$. 
Their commutators are  those of the \NGC algebra \bref{confalJM}.
In addition $\bCP_j$'s are no longer commuting but appears a central charge $Z=M$ \cite{Galajinsky:2011iz},
\be
[\bCP_j^a,\bCP_{k}^b]=-i\,\D^{ab}\,\D^{N,j+k}\,(-1)^{\frac{k-j+1}2}{k!\,j!}\,Z.
\label{centralextension}\ee
In appendix we show that in a non-linear realization method \cite{Coleman} the Lagrangian \bref{NderLag1} appears as the left invariant one-form associated with the central charge $Z$ in \bref{centralextension}. 
\vs

\section{Schrodinger Equation}
Now we consider the Schr\"odinger equation associated with the Hamiltonian \bref{hamiltonian}.
The canonical pairs $(\bX^j,\bP_j)$ satisfying \bref{Heisenberg} are realized as the hermitian 
linear operators $(\bx^j,-i\,\nabla_j)$ on complex wave functions $\psi(t,\bx^0,...,\bx^{\frac{N-1}2})$
with the inner product
\be \langle\psi_1|\psi_2\rangle=\int d^d\bx^0\cdots d^d\bx^{\frac{N-1}2}\, \,\ba{\psi_1(t,\bx^{j})}\,\psi_2(t,\bx^{j}),
\ee
where $\ba{\psi}$ is the complex conjugate function of ${\psi}$.
The Schr\"odinger equation $i\pa_t\psi=\CH\psi$ for the wave function 
$\psi(t,\bx^0,...,\bx^{\frac{N-1}2})$ is
\be
{{\Phi_S}}\, \psi(t,\bx^0,...,\bx^{\frac{N-1}2})=0. 
\label{NSchrodingereq}\ee
where ${{\Phi_S}}$ is the  Schr\"odinger differential operator defined by
\be
{{\Phi_S}}=i \frac{\pa}{\pa t}-\CH=i\frac{\pa}{\pa t}
+\left(i\sum_{j=0}^{\frac{N-3}2} \bx^{j+1}\nabla_j+\frac{1}{2\MN}\,(\nabla_{{\frac{N-1}2}})^2\right).
\label{NSchrodingerOp}\ee
The Schr\"odinger equation can be deduced from the action 
\bea\label{action}
I&=&\int dt\,d^d\bx^j\,{\ba{\psi(t,\bx^j)}}\,{{\Phi_S}}\,\psi(t,\bx^j),
\label{Schaction}\eea
which   is invariant under \NGC transformations 
in the scalar projective representation of the wave function. 
The \NGC algebra \bref{confalJM} is realized by the generators 
\bref{Canogene1} in the form of hermitian vector fields on the wave  
functions $\psi(t,\bx^1,...,\bx^{\frac{N-1}2})$, 
\bea
H&=&-\CH, \nn\\
D&=&-t\,\CH \,+
\sum_{j=0}^{\frac{N-1}2}\,(\frac{N}2-j)
\bx^j(-i\nabla_{j})-i\,d \frac{(N+1)^2}{16}, 
\nn\\
C&=&
-t^2\,\CH +\sum_{j=0}^{\frac{N-1}2}\,
\left((N-2j)t\,\bx^j+(N-j+1)j\,\bx^{j-1}\right)
\nn\\&&\times (-i\nabla_j) 
\,-\,{{\frac{\MN}2\,({\frac{N+1}2})^2(\bx^{{\frac{N-1}2}})^2}}{{-i\,d\,
\frac{(N+1)^2}{8}}}\,t, 
\nn\\
\bCP_j&=&j!\left( \sum_{\l=0}^j\,\frac{t^{j-\l}}{(j-\l)!}\,(-i\nabla_{\l})\right),\qquad (j=0,...,\frac{N-1}2),
\nn\\
\bCP_j&=&j!\left(  \sum_{\l=0}^{\frac{N-1}2}\,\frac{t^{j-\l}}{(j-\l)!}(-i\nabla_\l)-\MN \,\sum_{\l=\frac{N+1}2}^j
 (-)^{\frac{N+1}2+\l}
\right.\nn\\&&\left.\quad\times
\,\frac{t^{j-\l}}{(j-\l)!}\,\bx^{N-\l}\right),\qquad (j=\frac{N+1}2,...,N)
\nn\\Z&=&\MN,\label{Canogene2}\eea 
where $d$ is the spatial dimensions.
They are satisfying  the \NGC algebra \bref{confalJM} and \bref{centralextension}. 
These  generators commute with the ${{\Phi_S}}$, showing that they are constant of motion. Therefore the Schr\"odinger equation \bref{NSchrodingereq} remains 
invariant under the \NGC transformations. 
When $ \psi(t,\bx^j) $ is a solution of the Schr\"odinger equation then 
$\psi'(t,\bx^j)=e^{i\A G}\psi(t,\bx^j) $ is also a solution,
for $G= (H,D,C,\bCP_j,\bM_{ab})$.
Since the generators  are {hermitian}  the transformations $U=e^{i\A G}$ are 
unitary. The transformed wave function $\psi'(t,\bx^j)$ 
is also written in a form
\be
\psi'(t,\bx^j)=e^{i\A G} \psi(t,\bx^j)=e^{\CA+i\CB} \psi(t',\bx^{j'}),
\ee
using the  Schr\"odinger equation. 
Here ${\CA , \CB}$ are real functions of the coordinates and 
the transformation parameters $\A$'s.
$(t',\bx^{j'})$ are the coordinates transformed by the \NGC transformation. 

The $H$ transformation is time translation,  
\be
\psi'(t,\bx^j)=e^{iaH} \psi(t,\bx^j)
=e^{a\left(\pa_t \right)} \psi(t,\bx^j)= \psi(t+a,\bx^j)
\ee
For the {\it finite} scale transformation,  
\be
\psi'(t,\bx^j)=e^{i\lam D} \psi(t,\bx^j)
=\,e^{\lam\,d\,\frac{(N+1)^2}{16}}\,
\psi(e^\lam {t},e^{\lam(\frac{N}2-j)} {\bx^j}),
\ee
under which the action \bref{Schaction} is invariant. 
For the {\it finite} conformal transformation, $\A G=\kappa C$, 
\be
\psi'(t,\bx^j)=e^{i\kappa C} \psi(t,\bx^j)=e^{\CA+i\CB} \psi(t',\bx^{j'}). 
\ee
with
\bea
t'&=&\frac{t}{1-\kappa t},\quad 
\bx^{j'}=\sum_{r=0}^j\frac{{}_jC_r\,r!\kappa^r \,{}_{N-j+r}C_r}
{(1-\kappa t)^{N-2j+r}}\,\bx^{j-r},
\nn\\
e^\CA&=&(1-\kappa t)^{-\frac{d}2(\frac{N+1}{2})^2},\quad  
\CB=-{\frac{\kappa M}{2}}(\frac{N+1}{2})^2\,\frac{g_N}{(1-\kappa t)},
\nn\\
g_N&\equiv&\sum_{r,s=0}^{\frac{N-1}2}\frac{\gam_r\gam_s}{r!s!(r+s+1)}
(\frac{\kappa}{1-\kappa t})^{r+s}\,\bx^{\frac{N-1}2-r}\,\bx^{\frac{N-1}2-s},
\nn\\
&& \gamma_r\equiv{}_{\frac{N-1}{2}}C_{\frac{N-1}{2}-r}\,(r!)^2\,
{}_{\frac{N+1}{2}+r}C_{\frac{N+1}{2}}. \eea

For the {\it finite} $\bCP_j$ transformations,  
\be
\psi'(t,\bx^j)=
e^{i\sum_{j=0}^{N}\sbB^j\sbCP_j} \psi(t,\bx^j) = 
e^{-i2\pi\w_1(t,\bx;\sbB)}\,\psi(t,\bx^{j'}).\ee
The non-trivial projective phase \cite{bargmann} $\w_1$ associated with the $\bCP_j$ transformations is given by
\be
{2\pi}\w_1(t,\bx;\bB)={M}\sum_{k=0}^{\frac{N-1}2}
\,{(-)^{{\frac{N-1}2}+k}}(\bx^{k}+\frac{\7\bB^{k}}2)\7\bB^{N-k},
\ee
where the transformed coordinates are given by 
\be
{\bx^0}'= \bx^0+{\7\bB^0},\quad {\bx^\l}'= \bx^\l+\7\B^\l,
\quad {\7\bB^\l}(t) \equiv 
\frac{d^\l}{dt^\l}\sum_{j=0}^N\,t^{j}\bB^j.
\ee
We have a projective representation of the \NGC {group}. 

Under successive  $\bCP_j$ transformations  
{we get the non-trivial 2-cocycle $\w_2(\bB,\bB')$},
\be
U(\bB)U(\bB')
=e^{-i2\pi\w_2(\sbB,\sbB')}\,U(\bB+\bB'),
\ee
\be
2\pi\,\w_2(\bB,\bB')={M}\sum_{k=0}^{\frac{N-1}2}
\frac{(-)^{{\frac{N-1}2}+k}}2\left(\7\bB^{k}\,\7\bB^{'N-k}-
\7\bB^{'k}\,\7\bB^{N-k}\right).
\ee
The projective invariance of the Schr\"odinger equation 
is one to one correspondence with the fact 
that the higher order Lagrangian \bref{NderLag}, or the corresponding Lagrangian \bref{NderLag1} is invariant up to a total derivative under the  {\it finite }
$\bCP_j$ transformations \cite{levyleblond69}, 
\be
\CL_X'=\CL_X+\frac{d}{dt}(2\pi\,\w_1),
\ee
where transformations of $\bX^j$ and $\bY_j$ are , 
 \be
{\bX^\l}'= \bX^\l+
{\7\bB^\l}, \quad \bY_j'=\bY_j+ M(-)^{\frac{N-1}2+j}{\7\bB^{N-j}}.
\ee
These two  properties are  related to the fact that \NGC algebra has a central extension \bref{centralextension}  \cite{Marmo:1987rv}.

\vs
\section{Even $N$ Model in 2+1 Dimensions}
{In 2+1 dimensions we can  construct a local} higher order Lagrangian given by
\be\label{exotic}
\CL_X=\,\frac{\MN}{2}\,\ep^{ab}\,
\frac{d^{\frac{N}{2}}\bX_a}{dt^{\frac{N}{2}}}\frac{d^{\frac{N}{2}+1}\bX_b}{dt^{\frac{N}{2}+1}},
\ee
where $N$ is any positive even integer\footnote{The case N=2 was analyzed in \cite{Stichel:2003kh}.}.
The Lagrangian 
 equivalent to \bref{exotic} is  
\be
\CL_X=\frac{\MN}2\,\ep^{ab}\bX^{\frac{N}{2}}_a\,\dot\bX^{\frac{N}{2}}_b 
+ \sum_{j=0}^{\frac{N}2-1} \bY_j(\dot\bX^j-\bX^{j+1}).
\label{NderLageven}\ee
The Ostrogradsky momenta \cite{o}  are 
\be
\bP_j=\bY_j, \quad
 (j=0,...,\frac{N}2-1),\qquad \bP_{\frac{N}{2}}^b=\frac{\MN}2\,
\bX^{\frac{N}{2}}_a\ep^{ab}.
\ee
Using the second class constraints 
\be
\chi^a=\bP^a_{\frac N2}+\frac{\MN}2\ep^{ab}\bX^{\frac N2 b}=0,\qquad a=1,2,
\ee
the variables $\bX^{\frac N2}_a,(a=1,2)$ can be expressed in terms of the
independent $\bP^a_{\frac N2},(a=1,2)$ satisfying 
\be 
\left[\bP_{\frac{N}2}^a,\bP_{\frac{N}2}^b\right]=-i\frac{\MN}{4}\ep^{ab}.     
\ee
Note that the model has built in a noncommutative structure in phase space.
The Hamiltonian is 
\bea\label{hamiltonianeven}
\CH=\, \sum_{j=0}^{\frac{N}2-1} \bP_j\bX^{j+1}
\eea
and the equations of motion for $\bX\equiv \bX^0$ 
 gives $\frac{d^{N+1}}{dt^{N+1}}\bX^0=0.$ Note that in this case the order of derivatives, $(N+1),$ is odd. The canonical generators of \NGC algebra are 
\bea
H&=&-\CH,  
\nn\\
D&=&-t \,\CH +
\sum_{j=0}^{\frac{N}2-1}\,(\frac{N}2-j)
\frac{\{\bX^j, \bP_j\}_+}{2},\nn\\
C&=&-t^2\,\CH
+\sum_{j=0}^{\frac{N}2-1}\,
\left((N-2j)\,t\,\frac{\{\bX^j, \bP_j\}_+}{2} 
+(N-j+1)j\,\bX^{j-1}\,\bP_j\right)+N(\frac N2+1)\bP_{\frac{N}2}
\bX^{\frac{N}2-1},
\nn\\
\bCP_j&=&j!\left( \sum_{\l=0}^j\,\frac{t^{j-\l}}{(j-\l)!}\,\bP_\l\right),\qquad (j=0,...,\frac{N}2-1),
\nn\eea\bea
\bCP_j^b&=&j!\left(  \sum_{\l=0}^{\frac{N}2}\,\frac{t^{j-\l}}{(j-\l)!}\,\bP_\l^b-\MN\,\sum_{\l=\frac{N}2+1}^j\,  (-)^{\l}\times 
\,\frac{t^{j-\l}}{(j-\l)!}\,\bX^{N-\l}_a\ep^{ab}\right),\qquad 
(j=\frac{N}2,...,N),
\nn\\ {\bf J}_{ab}&=&\sum_{j=0}^{\frac{N}2}\,( \bX^j_{b}\,\bP_{ja}- \bX^j_{a}
\,\bP_{jb}),
\nn\\ {Z}&=&\MN. 
\label{Canogeneeven}\eea 
where the central charge $Z$  appears in 
\be [\bCP_j^a,\bCP_{k}^b]=-i\,\ep^{ab}\,\D^{N,j+k}\,(-1)^{\frac{j-k}2}{k!\,j!}\,Z.
\label{centexNeven}\ee
As in the odd $N$ case we can prove that these generators are constant of motion 
verifying the central extended \NGC algebra. 
We can also prove that the associated Schr\"odinger equation is invariant under a scalar projective representation of the \NGC group. 

\vs
\section{Summary and Discussion}
In this paper we have shown that the Schr\"odinger equation associated to the higher order non-relativistic particle is invariant under a projective representation of N(odd) GCA for any dimension. This results generalizes the well know result \cite{Niederer:1972zz} \cite{Hagen:1972pd}
that the ordinary Schr\"odinger equation is invariant under the Schr\"odinger (conformal)
group.  In the case of 2+1 dimensions we have seen that the exotic Schr\"odinger equation is invariant under N(even) conformal algebras. We expect these results could 
be useful among other areas in the non-relativistic conformal condensed matter correspondence \NGC/CMP.

The Hamiltonians \bref{hamiltonian} and \bref{hamiltonianeven} of present models are not positive semi-definite as is known in general
for higher time derivative Lagrangian systems\footnote{{A  fourth order gravity in (2+1) dimensions without  ghost has been studied in  \cite{arXiv:0904.4473}.}}.
In quantum theory generically the system will contain ghost degrees of freedom.
One possible procedure is to change the scalar product that has been applied to higher order harmonic oscillator \cite{Pais:1950za} in reference  \cite{arXiv:0706.0207}.
Another possibility is to eliminate the ghost spectrum by imposing a BRST like operator on the physical states \cite{DFTUZ-90-18}.


Possible extensions of the work is to consider the case of the \NGC algebra
for even $N$ in any dimensions.
We could also study {the symmetry properties of} the fourth order derivative harmonic oscillator \cite{Pais:1950za}, its generalizations.  We expect in this case we will have a realization of the Newton-Hooke \NGC algebra\cite{Galajinsky:2011iz}.
There are also possible higher order extensions of the Levy-Leblond equation \cite{LevyLeblond:1967zz} and the associated superconformal algebra \cite{Gauntlett:1990xq}.
\vs

\acknowledgments
 We thank Jos\'e de Azc\'arraga, Roberto Casalbuoni, Jaume Gomis, Jerzy Lukierski and Paul Townsend for comments.
We acknowledge partial financial support from projects FP2010-20807-C02-01,
2009SGR502 and CPAN Consolider CSD 2007-00042.

\appendix
\section{Particle Model action from Non-linear Realization}
Here we show how the NGC invariant action \bref{NderLag1}, for odd $N$,  is derived using the non-linear realization of the group G/H \cite{Coleman} (see also \cite{hep-th/0607057} and references there in), where G is the centrally extended algebra
  \bref{confalJM} with  \bref{centralextension}.
The left invariant MC form $\W=-ig^{-1}dg$ is expanded as
\be
\W=H\,L_H+D\,L_D+C\,L_C+\bP_j\,L_{\bP_j}+\frac12\bM_{ab}\,L_\bM^{ab}+Z\;L_Z
\label{MCW}\ee
and is satisfying the MC equation $d\W+i\W\wedge\W=0$. 
Using the NGC algebra \bref{confalJM} and  \bref{centralextension} their components satisfy the MC equations, 
\bea
dL_{H}&=&-L_{D}L_{H},\qquad dL_{D}=-2L_{C}L_{H},\qquad dL_{C}=L_{D}L_{C},\nn\\
dL_{\bP_j}^a&=&(j+1)\,L_{H}\,L_{\bP_{j+1}}^a+(j-\frac{N}2)\,L_{D}\,L_{\bP_j}^a+
(j-1-{N})\,L_{C}\,L_{\bP_{j-1}}^a-{L_{\bM}^a}_b L_{\bP_j}^b,\nn\\
d{L_{\bM}^a}_b&=&- {L_{\bM}^a}_c{L_{\bM}^c}_b,\quad dL_Z=\frac12\,\sum_{j=0}^{N}\, \frac{(-1)^j}{{}_NC_j} \, L_{\bP_j}\wedge L_{\bP_{N-j}}.
\label{MCeq}\eea
They are closed under $"d"$ that is equivalent with that the Jacobi identities 
of the algebra are satisfied. 
The right hand side of $ dL_Z$ is the WZ two form closed and invariant in the
non-extended algebra.

We parametrize the coset element as 
\be
g=e^{iHt}\,e^{i\bP_j\7\bX^j}e^{iC\s}\,e^{iD\rho}\,e^{iZc}\,.
\label{cosetnew}\ee
$(t,\7\bX^0)$ for the generators $(H,\bP_0)$ are 
identified to the $D=d+1$ dimensional space-time  coordinates.  
The left invariant MC form  components are 
\bea
L_H&=&e^{-\rho}\,dt,\qquad L_D=\,(d\rho -2\s\,dt),\qquad 
L_C=e^{\rho}\,(d\s+\s^2\,dt),
\label{LHDC}\\
L_{\bP_j}&=&e^{(j-\frac N2)\rho}\,\sum_{k=0}^{j}\,{}_{N-k}C_{j-k}(-\s)^{j-k}\,
(d\7\bX^k-(k+1)\7\bX^{k+1}\,dt), 
\label{LPnew}\nn\\
L_Z&=&
dc-\frac{\D^{ab}}2\sum_{j=0}^{N}\,\left(\frac{(-1)^j}{{}_NC_j}\,\7\bX^{N-j}_a \,d\7\bX^{j}_b+ \frac{(-1)^jj}{{}_NC_{j-1}}\,\7\bX^{N-j+1}_a \,\7\bX^{j}_b\,dt \right)
\label{LZoddNnew}\eea
Here and hereafter when there appear  $\7\bX^{-1}$ and $\7\bX^{N+1}$, they  are understood to be zero by definition. 
Note in the present parametrization of the coset 
$L_Z$ does not depend on neither $\rho$ nor $\s$. 

In the method of NLR the particle action is constructed from $\bM$ invariant one forms.
They are $ L_H,\, L_D,\,L_C,\, L_Z $
and we can use their linear combination as the invariant action. 
\be
I=
\int\,(\CL_c+\CL_\bX),
\quad
\CL_c=\left(b_H\,L_H+b_D\,L_D+b_C\,L_C\right)^*,\quad
\CL_\bX=\left(a\,L_Z\right)^*,
\ee
where * means pullback to the particle world line parametrized by $\tau$.
The first term $\CL_c$ depends only on the su(1,1) variables, $(t, \s, \rho),$
and giving one dimensional conformal mechanics Lagrangian \cite{deAlfaro:1977tj}.

We take the second term   $\CL_\bX$ as the particle Lagrangian
now depending on 
$t$ and $\7\bX^j$ in the present parametrization of the coset \bref{cosetnew}.
Using \bref{LZoddNnew} 
\bea
\CL_X&=&
a\left[\dot c-\frac{\D^{ab}}2\left(\sum_{j=0}^{N}\,\frac{(-1)^j}{{}_NC_j}\,\7\bX^{N-j}_a \,\dot{\7\bX}^{j}_b+ \sum_{j=1}^{N}\,\frac{(-1)^jj}{{}_NC_{j-1}}\,\7\bX^{N-j+1}_a \,\7\bX^{j}_b\,\dot t \right)\right]d\tau
\label{LZoddNnew2}\eea
and 
subtracting a surface term
it becomes
\bea
\CL_X 
&=&a\,\left[\sum_{j=0}^{\frac{N-3}{2}}\,\frac{(-1)^j(j+1)}{{}_NC_j}
 \,{\7\bX^{N-j}}\cdot
\left( \7\bX^{j+1}\dot t-\frac{1}{j+1}(\dot{\7\bX}^{j})
\right) \right.
\nn\\ 
&+&\left.\frac{(-1)^{{\frac{N-1}{2}}}({{\frac{N+1}{2}}}!)^2}{2N!\,\dot t}
\left( \7\bX^{{\frac{N+1}{2}}}\dot t-\frac{2}{N+1}\dot{\7\bX}^{\frac{N-1}{2}}
\right)^2
-\frac{(-1)^{{\frac{N-1}{2}}}({{\frac{N-1}{2}}}!)^2}{2N!\,\dot t}
\left(\dot{\7\bX}^{\frac{N-1}{2}}\right)^2\right]d\tau.\nn\\
\label{LLZNod} \eea
Here $\7\bX^{N-j}$ in the first term runs over 
$\7\bX^{\frac{N+3}{2}},...,\7\bX^{{N}} $ and they 
play roles of Lagrange multipliers giving their equations of motion ,
\bea
\7\bX^{j+1}=\frac{1}{j+1}(\frac{\dot{\7\bX}^{j}}{\dot t}),\qquad 
j=0,...,{\frac{N-3}{2}}.
\eea
$ \7\bX^{{\frac{N+1}{2}}}$ equation of motion determines $ \7\bX^{{\frac{N+1}{2}}}$ as  
\be
\7\bX^{{\frac{N+1}{2}}}=\frac{2}{N+1}(\frac{\dot{\7\bX}^{\frac{N-1}{2}}}{\dot t}).
\ee Using it back into the Lagrangian \bref{LLZNod} the second term can be
dropped. 
${\7\bX}^{{\frac{N+1}{2}}},...,{\7\bX}^1$ are 
solved iteratively in terms of ${\7\bX}^0$ and its derivatives,
\bea
  {\7\bX}^{j}&=&\frac{1}{j!}\,D^j{\7\bX}^{0},\qquad D\equiv \frac1{\dot t}
\frac{d}{d\tau},\qquad j=1,...,\frac{N+1}{2}. 
\eea
If we use them back in the Lagrangian we obtain 
\bea
\CL_X&=& 
a\,\frac{(-1)^{{\frac{N+1}{2}}}}{2N!}
(D^{\frac{N+1}{2}}{\7\bX}^{0})^2 \dot t\,d\tau
.\label{LLZNod2} \eea
In a static gauge $\dot t=1$ 
the Lagrangians \bref{LLZNod} and \bref{LLZNod2}  
become \bref{NderLag1} and \bref{NderLag} respectively by 
identifications  $ a=(-1)^{\frac{N+1}{2}}N!\,M $ and
\be
\7\bX^j=\frac{\bX^j}{j!},\,(j=0,...,\frac{N-1}{2}),\quad 
\7\bX^j=\frac{(-1)^{j}\,N!}{a\,j!}\bY^{N-j},\,(j=\frac{N+3}{2},...,N).
\ee
 

For even $N$ case the central extension is possible only in $D=1+2$ 
dimensions \bref{centexNeven}. Applying the similar discussions we 
arrive at the actions \bref{exotic} and \bref{NderLageven}.


\end{document}